\documentclass[aip,apl,reprint]{revtex4-1}
\usepackage{graphicx}
\usepackage{amsmath}
\usepackage{amssymb}
\usepackage{bm}
\usepackage{color}
\graphicspath{{figures_2/}}

\begin{document}

\title{Diffusive model of current-in-plane-tunneling in double magnetic tunnel junctions} 

\author{P.-Y. Clement}\email[Electronic mail: ]{pierre-yves.clement@cea.fr}\affiliation{SPINTEC, UMR CEA/CNRS/UJF-Grenoble 1/Grenoble-INP, INAC, Grenoble, F-38054, France}, 
\author{C. Ducruet}\affiliation{CROCUS-Technology, 5 place Robert Schuman, Grenoble, F-38054, France}, 
\author{C. Baraduc}\affiliation{SPINTEC, UMR CEA/CNRS/UJF-Grenoble 1/Grenoble-INP, INAC, Grenoble, F-38054, France}, 
\author{M. Chshiev}\affiliation{SPINTEC, UMR CEA/CNRS/UJF-Grenoble 1/Grenoble-INP, INAC, Grenoble, F-38054, France}, 
\author{B. Di\'{e}ny}\affiliation{SPINTEC, UMR CEA/CNRS/UJF-Grenoble 1/Grenoble-INP, INAC, Grenoble, F-38054, France}

\date{\today}

\begin{abstract}
We propose a model that describes current-in-plane tunneling transport in double barrier magnetic tunnel junctions in diffusive regime. Our study shows that specific features appear in double junctions that are described by introducing two typical length scales. The model may be used to measure the magnetoresistance and the resistance area product of both barriers in unpatterned stacks of double barrier magnetic tunnel junctions.
\end{abstract}


\maketitle 


Because of their applications in MRAM and hard disk drive read-heads, magnetic tunnel junctions have been extensively studied. In particular, double barrier magnetic tunnel junctions (DBMTJs) have been of particular interest due to high TMR ratios \cite{Jiang} and significantly slower TMR decay rates as a function of voltage, compared to single MTJs \cite{Montaigne}. Furthermore, it has been shown that, in such structures, spin transfer torque (STT) exerted on the magnetization of the central free layer can be enhanced \cite{brevet}, correlatively yielding a decrease in the critical current for STT magnetization switching \cite{Fuchs,Diao}. Finally, if the thickness of the central layer is small enough and its roughness sufficiently low, quantum well states may appear \cite{Kalitsov, Nozaki} and spin diode effect can be observed \cite{Iovan,Mair,Tiusan}. In this letter, we adress the case of DBMTJs developped for low consumption MRAMs where electrons are submitted to diffusive transport in all metallic layers.
 

The electrical transport in multilayered thin films having anisotropic in-plane and perpendicular-to-plane conductivities has been investigated in the context of metal/oxide multilayers \cite{Ernult}. Magnetic tunnel junctions constitute a particular case in this family of (metal/oxide) multilayered systems. For single barrier tunnel junctions, the barrier properties can be assessed just after deposition by measuring electrical transport in full sheet samples. Such measurements are performed with a multi-contact probe with various spacing between contacts. The resistance area product ($RA$) and magnetoresistance ratio ($MR$) can be extracted from the voltage variations versus probe position on the sample surface using current-in-plane-tunneling (CIPT) technique \cite{Worledge} implemented in Capres set-up. The CIPT technique leads to a significant gain of time since it allows assessing the good quality of the stack prior to microfabrication of the DBMTJ pillars. In this work, we developped an analytical model which allows extending this technique to double barrier diffusive stacks and provide a method to determine the junctions parameters for both junctions ($RA_{1}$, $RA_{2}$ and $MR_{1}$, $MR_{2}$) just after deposition. Our study shows that specific features appear in DBMTJs that are described by introducing two typical length scales.

\begin{figure}
\includegraphics[width=\columnwidth]{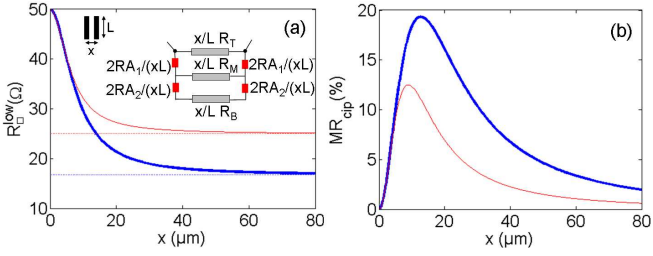}
\caption{Sheet resistance (a) and magnetoresistance (b) calculated in the toy model for single barrier (red, thin) and double barrier (blue, thick). $R_{T}=R_{M}=R_{B}=50\Omega/\square$; $RA_{1}=RA_{2}=1000\Omega\mu m^{2}$; $MR_{1}=MR_{2}=100\%$.}
\label{toy_model}
\end{figure}

We start with a simple description of double magnetic tunnel junctions as a network of resistors (toy model). It corresponds to the situation where two elongated contacts of length L separated by a distance $x$ are placed on the surface of the wafer. Ferromagnetic layers are modeled by
their sheet resistances $R_{T}$, $R_{B}$ and $R_{M}$ as well as the two barriers by their resistance area products $RA_{1}$ and $RA_{2}$. Thus, longitudinal conduction through each ferromagnetic layer is described by a resistance $R_{i}x/L$ (with $i=L,M,B$) while perpendicular conduction through the tunnel barriers is characterized by $RA_{j}/xL$ ($j=1,2$) (see inset of Fig.\ref{toy_model}). The equivalent resistance (Eq.(\ref{R_simple}) and (\ref{R'})) of the network is calculated by using basic electrokinetics.

\begin{equation}
\label{R_simple}
R_{tot}(x)=\frac{x}{L}\frac{R_{T}R'(x)}{R_{T}+R'(x)}\left(1+4\frac{R_{T}}{R'(x)}\frac{1}{4+\frac{x^{2}}{\lambda_{1}^{2}}\frac{R_{T}+R'(x)}{R_{T}+R_{M}}}\right)
\end{equation}
where $R'(x)$ is given by :
\begin{equation}
\label{R'}
R'(x)=\frac{R_{M}R_{B}}{R_{M}+R_{B}}\left(1+4\frac{R_{M}}{R_{B}}\frac{1}{4+\frac{x^{2}}{\lambda_{2}^{2}}}\right)
\end{equation}

with $\lambda_{1}^2=RA_{1}/(R_{T}+R_{M})$ and $\lambda_{2}^2=RA_{2}/(R_{B}+R_{M})$.

Fig.\ref{toy_model}a shows the sheet resistance for simple and double junctions. When the probes are close to each other (small $x$), simple and double barrier cases are equivalent. In this case, electrons travel only through the top layer. At large $x$, in contrast, the current flows through the whole structure, either in two or three layers in parallel: the calculated resistance is then equivalent to 2 or 3 sheet resistors in parallel. Let us now consider the magnetoresistance calculation (Fig.\ref{toy_model}b). A high (resp. low) value of $RA_{j}$ corresponds to an antiparallel (resp. parallel) configuration around the $j^{th}$ barrier. $MR_{cip}$ is defined as $MR_{cip} = 100(R_{high}-R_{low})/R_{low}$, considering that the magnetization of the two ferromagnetic external layers are pinned parallel and that only the central magnetization can switch. $MR_{cip}$ shows a maximum for a contact spacing that corresponds to the distance over which electrons must travel to reach the bottom layer. The maximum of $MR_{cip}$ in the double barrier case is therefore shifted towards larger $x$ values compared to the simple barrier case.

Even though the model gives relevant information about the transport in the double junction, it does not take into account the geometry of realistic contact probes. The exact problem of the current flow in double barrier stack connected at its top surface by two ponctual current probes can be solved considering the current probes as a source and a sink of electrons and applying the superposition theorem \cite{Worledge}. Since the probe spacing $x$ is always much larger than the layer thicknesses $t_{i}$ ($i=T,M,B$), one may assume that the voltage drop in the vertical direction only appears across the barriers. Current conservation is applied to an infinitesimal cylinder around the current probe and then to a shell between $r$ and $r+dr$:

\begin{equation}
\label{conservation0}
I=2\pi r \left( J_{T}(r)t_{T}+J_{M}(r)t_{M}+J_{B}(r)t_{B} \right)
\end{equation}

\begin{equation}
\label{conservation1}
J_{Z}^{(1)}(r)+t_{T} \frac{\partial J_{T}(r)}{\partial r} + \frac{1}{r} J_{T}(r)t_{T}=0
\end{equation}
\begin{equation}
\label{conservation2}
J_{Z}^{(2)}(r)-J_{Z}^{(1)}(r)+t_{M}\frac{\partial J_{M}(r)}{\partial r}+\frac{1}{r} J_{M}(r)t_{M}=0
\end{equation}

where $J_{i}$, $i=T, M, B$ are the longitudinal current densities through the ferromagnetic layers and $J_{Z}^{(j)}$, $j=1$ or $2$ are the current densities across the barriers. We finally apply the mesh rule to the loop around the top and bottom barriers:

\begin{equation}
\label{conservation3}
R_{T}J_{T}(r)t_{T}-R_{M}J_{M}(r)t_{M}+RA_{1}\frac{\partial J_{Z}^{(1)}(r)}{\partial r}=0
\end{equation}
\begin{equation}
\label{conservation4}
R_{M}J_{M}(r)t_{M}-R_{B}J_{B}(r)t_{B}+RA_{2}\frac{\partial J_{Z}^{(2)}(r)}{\partial r}=0
\end{equation}

By combining equations (\ref{conservation0}) to (\ref{conservation4}), we get the following fourth-order differential equation.

\begin{widetext}
\begin{equation}
\begin{split}
\label{EDET}
\frac{\partial ^{4}E_{T}(r)}{\partial r^{4}}+\frac{2}{r}\frac{\partial ^{3}E_{T}(r)}{\partial r^{3}}-\left(\frac{3}{r^{2}}+\frac{1}{\overline{\lambda}^{2}}\right)\frac{\partial ^{2}E_{T}(r)}{\partial r^{2}}+\frac{1}{r}\left(\frac{3}{r^{2}}-\frac{1}{\overline{\lambda}^{2}}\right)\frac{\partial E_{T}(r)}{\partial r} \\ +\left[\frac{1}{r^{2}}\left(-\frac{3}{r^{2}}  +\frac{1}{\overline{\lambda}^{2}}\right)+\frac{1}{\Lambda^{4}}\right]E_{T}=\frac{R_{T}R_{M}R_{B}}{RA_{1}RA_{2}}\frac{I}{2\pi r}
\end{split}
\end{equation}
\end{widetext}

where $E_{T}(r)=t_{T}R_{T}J_{T}(r)$ has the dimension of an electric field and $\frac{1}{\overline{\lambda}^{2}}= \frac{1}{\lambda_{1}^{2}}+\frac{1}{\lambda_{2}^{2}}$, $\Lambda^{4}=\frac{RA_{1}RA_{2}}{R_{T}R_{M}+R_{M}R_{B}+R_{B}R_{T}}$

As the double tunnel junction problem can be seen as the interweaving of two simple junction ones, one can successively apply the differential expression found in the simple junction case \cite{Worledge}.
\begin{equation}
\label{epsilon}
\epsilon_{T}(r)=\frac{\partial ^{2}E_{T}(r)}{\partial
r^{2}}+\frac{1}{r}\frac{\partial E_{T}(r)}{\partial
r}-\left(\frac{1}{r^{2}}+\frac{1}{\lambda_{\pm}^{2}}\right)E_{T}(r)
\end{equation}
\begin{equation}
\label{EDepsilon}
\frac{\partial ^{2} \epsilon_{T}(r)}{\partial
r^{2}}+\frac{1}{r}\frac{\partial \epsilon_{T}(r)}{\partial
r}-(\frac{1}{r^{2}}+\frac{1}{\lambda_{\mp}^{2}})\epsilon_{T}(r)
\end{equation}

Eq.(\ref{EDepsilon}) exactly gives the left member of Eq.(\ref{EDET}) provided that $\lambda_{\pm}=\left[\frac{\Lambda^{4}}{2}\left(\frac{1}{\overline{\lambda}^{2}}\pm\sqrt{\frac{1}{\overline{\lambda}^{4}}-\frac{4}{\Lambda^{4}}}\right)\right]^{\frac{1}{2}}$. Thus the problem can be solved by successively integrating two well-known second-order differential equations. The voltage drop is then calculated by integrating the electric field $E_{T}$ between the two central probes. If we consider four equally spaced probes, we obtain the sheet resistance \cite{vanderpauw} given by $R_{\square}=\frac{\pi}{\ln(2)}R$:

\begin{widetext}
\begin{equation}
\label{Rsquare}
R_{\square}(x)=R_{+}\left(K_{0}\left(\frac{x}{\lambda_{+}}\right)-K_{0}\left(\frac{2x}{\lambda_{+}}\right)\right)+R_{-}\left(K_{0}\left(\frac{x}{\lambda_{-}}\right)-K_{0}\left(\frac{2x}{\lambda_{-}}\right)\right)+\overline{R}
\end{equation}
\end{widetext}

where $K_{0}$ is the Bessel function of the second kind of order zero, $\overline{R}=\frac{R_{T}R_{M}R_{B}}{R_{T}R_{M}+R_{M}R_{B}+R_{T}R_{B}}$ is the sheet resistance of the three ferromagnetic layers in parallel, and $R_{\pm}$ is given by:
\begin{equation}
R_{\pm}=\frac{\lambda_{\pm}}{\ln(2)}\frac{R_{T}\left(1-\frac{R_{T}}{R_{T}+R_{M}}\left(\frac{\lambda_{\mp}}{\lambda_{1}}\right)^{2}\right)-\overline{R}}{\lambda_{\pm}^{2}-\lambda_{\mp}^{2}}
\end{equation}

\begin{figure}
\includegraphics[width=\columnwidth]{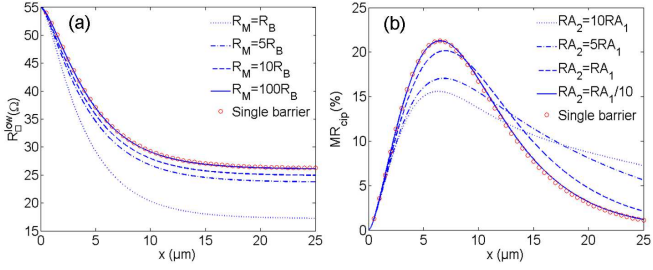}
\caption{(a) CIPT resistance for various $R_{M}$ with $R_{T}=55\Omega/\square$, $R_{B}=50\Omega/\square$ and $RA_{1}=RA_{2}=1000\Omega\mu m^{2}$. (b) CIPT magnetoresistance for various $RA_{2}$ with $RA_{1}=1000\Omega\mu m^{2}$, $R_{T}=R_{M}=R_{B}=50\Omega/\square$ and $MR_{1}=MR_{2}=100\%$. For (a) and (b), the red circles correspond to the single barrier model.}
\label{validation}
\end{figure}

Compared to the single barrier case \cite{Worledge}, we notice that there are now four Bessel functions characterized by two different length scales $\lambda_{+}$ and $\lambda_{-}$. In the following, an interpretation of these two length scales is proposed. Nevertheless it is crucial to first validate Eq.(\ref{Rsquare}) by considering some limit cases. For large probe spacing, we recover the sheet resistance $\overline{R}$ of the three layers in parallel since $K_{0}(x)-K_{0}(2x)$ rapidly converges to zero for $x \geq 5$. For small probe spacing, the current flows only through the top layer. Considering that $\lim\limits_{x \to 0}K_{0}(x)-K_{0}(2x)=\ln(2)$ and $R_{+}+R_{-}=(R_{T}-\overline{R})/\ln(2)$, Eq.(\ref{Rsquare}) exactly gives the expected result $R_{\square}=R_{T}$. Let us further check our model by considering two specific situations. If the width of the central magnetic layer tends to zero ($ie$ $R_{M}\rightarrow+\infty$), the two barriers become closer and closer until forming a single barrier junction with $RA_{eff}=RA_{1}+RA_{2}$ (Fig.\ref{validation}a). Second, by reducing $RA_{2}$ to zero, the structure also converges toward the single barrier case with a bottom layer thickness equal to $t_{M}+t_{B}$ (Fig.\ref{validation}b). Using these two examples, we checked the validity of the model by recovering the simple junction curves (red circles in Fig.\ref{validation}). 

\begin{figure}
\includegraphics[width=\columnwidth]{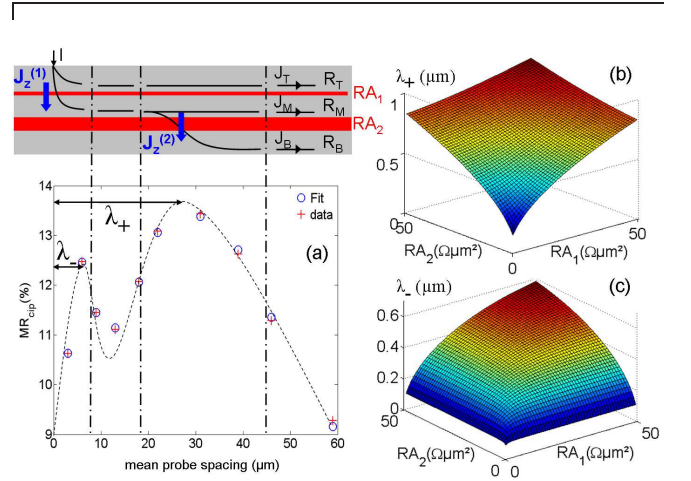}
\caption{(a) Comparison of Capres data with our model using $RA_{1}=700\Omega.\mu m^{2}$, $RA_{2}=12.6 k\Omega.\mu m^{2}$, $MR_{1}=160\%$, $MR_{2}=50\%$, $R_{T}=36\Omega/\square$, $R_{M}=69\Omega/\square$, $R_{B}=13\Omega/\square$ (the dashed line is a guide for the eye); on top a sketch of the current flow (black) through the structure compared to the two length scales $\lambda_{+}$ and $\lambda_{-}$; (b), (c) $\lambda_{+}$ and $\lambda_{-}$ as a function of $RA_{1}$ and $RA_{2}$.}
\label{interpretation}
\end{figure}

Let us now discuss the physical meaning of the two length scales $\lambda_{+}$ and $\lambda_{-}$ by considering a situation where the top barrier has a much lower resistance than the bottom one ($RA_{1}<<RA_{2}$). For that purpose, a DBMTJ was deposited by sputtering with the following composition: Ta 5/Ru 7/Ta 5/PtMn 20/CoFe 2/Ru 0.8/CoFeB 2/MgO 3.3/CoFeB 20/MgO 2.2/CoFeB 2/NiFe 3/FeMn 12/Ru 5 (nm). Two successive annealings under magnetic field at 300$^{\circ}$C and 180$^{\circ}$C align in parallel the magnetizations of the layers above and below the free layer. Then a small magnetic field in the opposite direction (during Capres measurement) switches the middle free layer in order to obtain the antiparallel configuration. Capres measurement of the $MR_{cip}$ shows an original trend (Fig.\ref{interpretation}a), that is perfectly fitted by our model. There are now two maxima, each of them related to a characteristic length scale : the first maximum (at small $x$) is controlled by $\lambda_{-}$ while the second one is governed by $\lambda_{+}$. Thus both characteristic lengths can be interpreted as sketched in Fig.\ref{interpretation}. Electrons cross the top tunnel barrier on a length scale equal to $\lambda_{-}$, leading thereby to a magnetoresistance maximum. When $x$ becomes larger, $MR_{cip}$ decreases since the current flows mostly in parallel through the two upper ferromagnetic layers. The second maximum corresponds to the distance $\lambda_{+}$ at which electrons tunnel through the second barrier. Finally, at large $x$, $MR_{cip}$ goes back to zero since current flows through three ferromagnetic layers in parallel.

Dependances of $\lambda_{+}$ and $\lambda_{-}$ as a function of both $RA_{1}$ and $RA_{2}$ are given in Fig.\ref{interpretation}b and c. We primarily notice that both $\lambda_{-}$ and $\lambda_{+}$ are symmetric with respect to $RA_{1}$ and $RA_{2}$. Moreover both length scales increase with $RA_{1}$ and $RA_{2}$. However the surface shapes representing $\lambda_{-}$ and $\lambda_{+}$  are clearly different. $\lambda_{+}$ is large when only one barrier has a large $RA$ value, thus indicating that the transport is governed by the thickest barrier. Moreover $\lambda_{+}$ becomes even larger when both $RA_{1}$ and $RA_{2}$ increase. These observations are consistent with our interpretation of $\lambda_{+}$ as the length over which electrons travel through the whole structure. In contrast to $\lambda_{+}$, the length scale $\lambda_{-}$ evolves quite differently. It stays small as long as at least one barrier has a small $RA$ value, thus leading us to interpret $\lambda_{-}$ as the characteristic length scale for transport through the thinnest barrier of the DBMTJ. Consistently with this interpretation, we observed that $\lambda_{+}=2\lambda_{-}$ along the symmetry line $RA_{1}$=$RA_{2}$.

In conclusion, we have developed an analytical model that describes in-plane diffusive transport in DBMTJs and shown that these structures may present original features such as a magnetoresistance with two maxima as a function of probe spacing. Our results are interpreted by introducing two length scales $\lambda_{+}$ and $\lambda_{-}$. Finally, this model can be used to extract the four fundamental characteristics of DBMTJs ($RA_{1}$, $RA_{2}$, $MR_{1}$ and $MR_{2}$) by implementing our fitting procedure on Capres set-up.

\begin{acknowledgments}
This work was partially funded by the European Commission through the ERC Adv Grant HYMAGINE No. 246942 
\end{acknowledgments}

%
\end{document}